\documentclass[journal]{IEEEtran}

\makeatletter
\long\def\@makecaption#1#2{\ifx\@IEEEtablestring%
\footnotesize{\normalfont\footnotesize #1}\\
{\normalfont\footnotesize\scshape #2}%
\@IEEEtablecaptionsepspace
\else
\@IEEEfigurecaptionsepspace
\setbox\@tempboxa\hbox{\normalfont\footnotesize {#1.}~~ #2}%
\ifdim \wd\@tempboxa >\hsize%
\setbox\@tempboxa\hbox{\normalfont\footnotesize {#1.}~~ }%
\parbox[t]{\hsize}{\normalfont\footnotesize \noindent\unhbox\@tempboxa#2}%
\else
\hbox to\hsize{\normalfont\footnotesize\hfil\box\@tempboxa\hfil}\fi\fi}
\makeatother
\def\tablename{Tab.}

\usepackage{color}
\usepackage[pdftex]{graphicx}
\usepackage[cmex10]{amsmath}
\usepackage{amsfonts}
\usepackage{cite}
\usepackage{url}
\usepackage{hyperref}
\usepackage{tikz}
\usepackage[normalem]{ulem}

\hypersetup{
	colorlinks=true, 
	linkcolor=blue!70!black, 
	citecolor=red!70!black, 
	filecolor=blue!70!black, 
	urlcolor=blue!70!black, 
	backref=true
}

\newcommand{\J}{\mathrm{j}}                 

\providecommand{\D}{\,\mathrm{d}}           
\providecommand{\V}[1]{\boldsymbol{#1}}     
\providecommand{\M}[1]{\mathbf{#1}}         
\newcommand{\ABS}[1]{\left| #1 \right|}     

\providecommand{\herm}{\mathrm{H}} 
\providecommand{\trans}{\mathrm{T}}

\DeclareMathOperator*{\argmin}{arg\,min}

\providecommand{\srcRegion}{\varOmega} 
\providecommand{\Ivec}{\M{I}}

\providecommand{\ZVAC}{Z_0}  
\providecommand{\Prad}{P_\mathrm{r}}

\providecommand{\We}{W_\mathrm{e}}
\providecommand{\Wm}{W_\mathrm{m}}
\providecommand{\Xe}{\M{X}_\mathrm{e}}
\providecommand{\Xm}{\M{X}_\mathrm{m}}

\providecommand{\gXe}{\widetilde{\M{X}}_\mathrm{e}}
\providecommand{\gXm}{\widetilde{\M{X}}_\mathrm{m}}
\providecommand{\genRop}{\widetilde{\M{R}}}  
\providecommand{\Qmin}{Q_\mathrm{lb}}  
\providecommand{\Qdgmin}{\widetilde{Q}_\mathrm{lb}}  
\providecommand{\QE}{Q_\mathrm{e}}  
\providecommand{\QM}{Q_\mathrm{m}}  
\providecommand{\spar}{\nu}  
\providecommand{\Qpar}{\widetilde{Q}_\spar}  
\providecommand{\QEpar}{Q_{\mathrm{e}\spar}}
\providecommand{\QMpar}{Q_{\mathrm{m}\spar}}

\providecommand{\rvh}{\hat{\V{r}}} 
\providecommand{\evh}{\hat{\V{e}}} 
\providecommand{\thetavh}{\hat{\V{\theta}}} 
\providecommand{\phivh}{\hat{\V{\phi}}} 
\providecommand{\sphi}{\alpha}  

\providecommand{\parpar}{\varsigma}

\providecommand{\mixModeAmp}{\chi}
\providecommand{\mixModePhase}{\phi}  

\newcommand{\cStructA}{black!10!white}
\newcommand{\cStructB}{black!30!white}

\newcommand{\ie}{\textit{i.e.}{}}
\newcommand{\eg}{\textit{e.g.}{}}
\newcommand{\cf}{\textit{cf.}{}}

\begin{document}
\title{Minimization of Antenna Quality Factor}
\author{Miloslav~Capek,~\IEEEmembership{Member,~IEEE,}
	    Mats~Gustafsson,~\IEEEmembership{Member,~IEEE,}
	    Kurt~Schab,~\IEEEmembership{Member,~IEEE}
\thanks{Manuscript received  \today; revised \today.
This work was supported by the Czech Science Foundation under project No. 15-10280Y and by the Swedish Foundation for
Strategic Research (SSF) under the program Applied Mathematics and the project Complex analysis and convex optimization for EM design.  Kurt Schab was supported by the Intelligence Community Postdoctoral Research Fellowship Program.  All statements of fact, opinion, or analysis expressed are those of the author and do not reflect the official positions or views of the Intelligence Community or any other U.S. Government agency.  Nothing in the contents should be construed as asserting or implying U.S. Government authentication of information or Intelligence Community endorsement of the author’s views.}
\thanks{M.~Capek is with the Department of Electromagnetic Field, Faculty of Electrical Engineering, Czech Technical University in Prague, 166~27 Prague, Czech Republic
(e-mail: miloslav.capek@fel.cvut.cz).}
\thanks{M.~Gustafsson is with the Department of Electrical and Information Technology,
Lund University, 221~00 Lund, Sweden (e-mail: mats.gustafsson@eit.lth.se).}
\thanks{K.~Schab is with the Department of Electrical and Computer
Engineering, Antennas and Electromagnetics Laboratory, North Carolina State University, Raleigh, NC, USA (e-mail: krschab@ncsu.edu).}
}

\markboth{Journal of \LaTeX\ Class Files,~Vol.~XX, No.~XX, \today}
{Capek, Gustafsson, Schab: Minimization of Antenna Quality Factor}
\maketitle

\begin{abstract}
The optimal currents on arbitrarily shaped radiators with respect to the minimum quality factor~$Q$ are found using a simple and efficient procedure. The solution starts with a reformulation of the problem of minimizing quality factor~$Q$ as an alternative, so-called dual, problem. Taking advantage of modal decomposition and group theory, it is shown that the dual problem can easily be solved and always results in minimal quality factor~$Q$. Moreover, the optimization procedure is generalized to minimize quality factor~$Q$ for embedded antennas, with respect to the arbitrarily weighted radiation patterns, or with prescribed magnitude of the electric and magnetic near-fields. The obtained numerical results are compatible with previous results based on composition of modal currents, convex optimization, and quasi-static approximations; however, using the methodology in this paper, the class of solvable problems is significantly extended.
\end{abstract}


\IEEEpeerreviewmaketitle

\section{Introduction}
\label{sec:Intro}
\IEEEPARstart{T}{he} quality factor~$Q$ (or Q-factor) is of primary importance in the design and analysis of electrically small antennas because of its approximate proportionality to fractional bandwidth \cite{VolakisChenFujimoto_SmallAntennas,Hansen_ESA_SuperDirAndSuperconducAntennas,GustafssonTayliCismasu_PhysicalBoundsOfAntennasBook}. Classical approaches to estimate the value of an antenna's quality factor~$Q$ center around evaluating the frequency sensitivity of the antenna in different ways: from the antenna's electric field integral equation \cite{Vandenbosch_ReactiveEnergiesImpedanceAndQFactorOfRadiatingStructures}, equivalently from its impedance (reactance) matrix~\cite{Gustafsson_OptimalAntennaCurrentsForQsuperdirectivityAndRP}, or directly from its input impedance~\cite{YaghjianBest_ImpedanceBandwidthAndQOfAntennas,GustafssonJonsson_AntennaQandStoredEnergiesFieldsCurrentsInputImpedance}. In many cases, these techniques can reasonably predict the bandwidth potential for high-$Q$ antennas~\cite{VolakisChenFujimoto_SmallAntennas,CapekJelinekHazdra_OnTheFunctionalRelationBetweenQfactorAndFBW,GustafssonJonsson_AntennaQandStoredEnergiesFieldsCurrentsInputImpedance}.

The determination of the fundamental bounds on antenna performance has evolved from methods studying canonical bodies, such as spherical shells~\cite{Chu_PhysicalLimitationsOfOmniDirectAntennas}, to the more general analyses of arbitrarily shaped radiators~\cite{GustafssonTayliCismasu_PhysicalBoundsOfAntennasBook}. Much of this work is motivated by seeking an optimal radiator's performance quantified in terms of quality factor~$Q$, antenna gain $G$, and radiation efficiency $\eta_\mathrm{rad}$ as functions of electrical size~\cite{VolakisChenFujimoto_SmallAntennas,Hansen_ESA_SuperDirAndSuperconducAntennas}. Certain antenna parameters (or their combinations) can be analyzed in the static limit~\cite{Yaghjian_MinimumQforLossyAndLosslessESA} or be reformulated as convex optimization problems for larger structures~\cite{BoydVandenberghe_ConvexOptimization,GustafssonTayliEhrenborgEtAl_AntennaCurrentOptimizationUsingMatlabAndCVX}. Effects of finite ground planes, lossy materials, or magnetic currents can also be included~\cite{TayliGustafsson_PhysicalBoundsForAntennasAboveGND, Gustafsson_OptimalAntennaCurrentsForQsuperdirectivityAndRP,GustafssonTayliEhrenborgEtAl_AntennaCurrentOptimizationUsingMatlabAndCVX}. However, the unconstrained minimization of quality factor~$Q$ has not been formulated as a convex problem, making its solution much more difficult.

The method employed in this paper proposes a dual problem formulation or, in other words, a relaxation of the unconstrained optimization problem~\cite{GustafssonTayliEhrenborgEtAl_AntennaCurrentOptimizationUsingMatlabAndCVX}. Determining the optimal perturbation parameter while taking the smallest eigenvalue of the related generalized eigenvalue problem forms a problem which is directly solvable via the bisection algorithm~\cite{NocedalWright_NumericalOptimization}. Because the original problem of minimizing quality factor~$Q$ is generally non-convex, the difference between the dual solution and the solution to the original problem (\ie, the duality gap) can be non-zero~\cite{BoydVandenberghe_ConvexOptimization,GustafssonTayliEhrenborgEtAl_AntennaCurrentOptimizationUsingMatlabAndCVX}. However, we show in this paper that through proper treatment of degenerate eigenmodes within the dual problem the duality gap can always be made zero.  Taking advantage of the von Neumann-Wigner theorem~\cite{vonNeumannWigner_OnTheBehaviourOfEigenvaluesENG} and work on crossing avoidance~\cite{SchabEtAl_EigenvalueCrossingAvoidanceInCM} and group theory~\cite{Knorr_1973_TCM_symmetry,SchabBernhard_GroupTheoryForCMA} we also determine when such degeneracies occur based on symmetry.

The presented approach is generalized to several related but previously unsolved antenna problems. The minimum quality factor~$Q$ for antennas near finite ground planes is found by dividing the antenna body into controllable and uncontrollable sub-regions~\cite{Gustafsson_OptimalAntennaCurrentsForQsuperdirectivityAndRP}. It is also shown that minimization of the stored energy with arbitrary weighted constraints on the radiation intensity and the power in the electric and magnetic near-fields can be solved in similar fashion. This gives physical bounds for several antenna problems. Particularly, usage of the radiation intensity in a single direction generalizes the bounds on the partial gain in~\cite{Gustafsson_OptimalAntennaCurrentsForQsuperdirectivityAndRP,GustafssonTayliCismasu_PhysicalBoundsOfAntennasBook} to the gain.  

The proposed minimization procedure is computationally inexpensive and easy to implement. The results are also comparable with recently published methods using modal decomposition~\cite{ChalasSertelVolakis2016EarlyAccess, CapekJelinek_OptimalCompositionOfModalCurrentsQ,JelinekCapek_OptimalCurrentsOnArbitrarilyShapedSurfaces} to determine the optimal currents with respect to the minimum quality factor~$Q$. For simplicity, the presented analysis is restricted to compactly supported electric surface current densities in free space. The results can directly be generalized to volumetric and magnetic current densities~\cite{JonssonGustafsson_StoredEnergiesInElectricAndMagneticCurrentDensities_RoyA,Kim_LowerBoundsOnQForFinizeSizeAntennasOfArbitraryShape}.

The paper is organized as follows. The optimization procedure used to determine the minimum quality factor~$Q$ is introduced in Section~\ref{sec:Method} and its important properties are discussed in Section~\ref{sec:Explanation}, including a proof of method's convergence to the global minimum. The method is generalized in Section~\ref{sec:Generalization} towards the utilization of advanced constraints. Validity of the method is verified by examples of various complexity in Section~\ref{sec:Res}. The consequences of the method are discussed in Section~\ref{sec:Disc} and the paper is concluded in Section~\ref{sec:Concl}.


\section{Minimization of Quality Factor~$Q$}
\label{sec:Method}

In this section, the problem of minimizing quality factor~$Q$ is formulated, along with the associated dual problem and its solution.

\subsection{Problem Statement}
\label{sec:Method:OrigProblem}

For a given surface or volume of compact support $\srcRegion$, we would like to determine the minimal quality factor~$\Qmin$, defined as
\begin{equation}
\Qmin = \min_\Ivec \left\{ Q \left( \Ivec \right) \right\},
\label{eq:QminDef}
\end{equation}
in which $\Ivec$ is yet unknown vector containing expansion coefficients of the current density \mbox{$\V{J}\left( \V{r} \right)$}, \mbox{$\V{r} \in \srcRegion$}, represented in a suitable basis $\left\{\V{\psi}_n\right\}$ \cite{Harrington_FieldComputationByMoM, ChewTongHu_IntegralEquationMethodsForElectromagneticAndElasticWaves}, \ie,
\begin{equation}
\V{J} \left( \V{r}\right) = \sum\limits_{n=1}^N I_n \V{\psi}_n \left( \V{r}\right).
\label{eq:IvecDef}
\end{equation}
The quality factor~$Q$ in \eqref{eq:QminDef} is defined as \cite{IEEEStd_antennas}
\begin{equation}
Q \left( \Ivec \right) = \frac{2 \omega \max\left\{\Wm, \We \right\}}{\Prad} \equiv \max \left\{ \QM, \QE \right\}.
\label{eq:Qdef}
\end{equation}
For a given current distribution represented by $\Ivec$, the stored magnetic and electric energies are approximated as \cite{Gustafsson_OptimalAntennaCurrentsForQsuperdirectivityAndRP, Vandenbosch_ReactiveEnergiesImpedanceAndQFactorOfRadiatingStructures}
\begin{subequations}
\label{eq:WDef}
\begin{equation}
	\Wm \approx \frac{1}{8}\Ivec^\herm \left( \frac{\partial\M{X}}{\partial\omega} + \frac{\M{X}}{\omega} \right)\Ivec = \frac{1} {4\omega}\Ivec^\herm\Xm\Ivec,
\label{eq:WmDef}
\end{equation}
\begin{equation}
	\We \approx \frac{1}{8}\Ivec^\herm \left( \frac{\partial\M{X}}{\partial\omega} - \frac{\M{X}}{\omega} \right)\Ivec = \frac{1} {4\omega}\Ivec^\herm\Xe\Ivec,
\label{eq:WeDef}
\end{equation}
\end{subequations}
respectively, and the radiated power $\Prad$ is given by
\begin{equation}
	\Prad = \frac{1}{2} \Ivec^\herm \M{R} \Ivec.
\label{eq:PrDef}
\end{equation}
Here, \mbox{$\M{Z} = \M{R} + \J \left(\Xm - \Xe\right)$} is the impedance matrix based on the Galerkin method \cite{Harrington_FieldComputationByMoM}, \mbox{$\M{X} = \Xm - \Xe$}, $\omega$ is angular frequency, and superscript $^\herm$ denotes the complex conjugate transpose. The quadratic forms~\eqref{eq:WmDef} and~\eqref{eq:WeDef} based on the stored energy matrices $\Xm$ and $\Xe$ accurately predict the stored energies $\Wm$ and $\We$ for currents with support up to the electrical size of at least \mbox{$ka = 1$}, in which $k$ is the wavenumber and $a$ is the radius of smallest sphere circumscribing~$\srcRegion$~\cite{GustafssonTayliEhrenborgEtAl_AntennaCurrentOptimizationUsingMatlabAndCVX}.  These matrices can easily be obtained from method-of-moments codes \cite{CismasuGustafsson_FBWbySimpleFreuqSimulation}.

\subsection{Formulation and Solution of Dual Problem}
\label{sec:Method:DualProblem}
 
Following \cite{GustafssonTayliEhrenborgEtAl_AntennaCurrentOptimizationUsingMatlabAndCVX}, the original problem in \eqref{eq:QminDef} can be relaxed by convex combination\footnote{This way, the original non-convex problem is rewritten as a convex optimization problem \cite{BoydVandenberghe_ConvexOptimization}.} of $\QM$ and $\QE$ from \eqref{eq:Qdef}, \ie, for any current $\Ivec$,
\begin{equation}
\max \left\{ \QM, \QE \right\} \geq \left(1 - \spar \right) \QM + \spar \QE
\label{eq:QnuDef}
\end{equation}
for \mbox{$0 \leq \spar \leq 1$}.
Minimizing both sides of \eqref{eq:QnuDef} over $\Ivec$ gives the lower bound $\Qmin$ in \eqref{eq:QminDef} on the LHS, and the solution to the dual problem for a given value of the perturbation parameter, $\spar$, on the RHS of \eqref{eq:QnuDef}, \ie,
\begin{equation}
\Qmin \geq \min_\Ivec \left\{\left(1 - \spar \right) \QM + \spar \QE \right\}.
\label{eq:dualproblemDef}
\end{equation}
We denote the solution to the RHS of \eqref{eq:dualproblemDef} as $\Qpar$ and
\begin{equation}
\Ivec_\spar = \argmin_\Ivec \left\{\left(1 - \spar \right) \QM + \spar \QE \right\}
\end{equation}
which can be easily obtained by finding the smallest eigenvalue solution to 
\begin{equation}
\big( \left(1 - \spar \right) \Xm + \spar \Xe \big) \Ivec_\spar = \Qpar \M{R} \Ivec_\spar.
\label{eq:eigenvalueDef}
\end{equation}
Inserting the solution $\Qpar$ into the RHS of \eqref{eq:dualproblemDef} and recognizing that \mbox{$Q \left( \Ivec_\spar \right)$} from \eqref{eq:Qdef} must be greater than or equal to the lower bound $\Qmin$, we arrive at the three-part inequality
\begin{equation}
\label{eq:QpartsDef}
\max \{ \QMpar, \QEpar \} \geq \Qmin \geq \Qpar, 
\end{equation}
where the following notation is adopted:
\begin{equation}
\label{eq:QmQeDef}
\QMpar = \QM\left( \Ivec_\spar \right) \quad \textrm{and} \quad \QEpar = \QE\left(  \Ivec_\spar \right).
\end{equation}
Using these quantities, $\Qpar$ is also written as a convex combination
\begin{equation}
\label{eq:QnuTildeDef}
\Qpar = \left( 1 - \spar \right)\QMpar + \spar\QEpar.
\end{equation}
If equality holds between the far-left and far-right sides of~\eqref{eq:QpartsDef}, then the duality gap is zero and $\Qmin$ is effectively found by solution of the dual problem. In the following section, we show that this is always the case. Hence, the concave dual problem of maximizing $\Qpar$ over $\spar$ can always be used to find the solution to the unconstrained minimization problem, \ie,
\begin{equation}
\label{eq:maxProblemDef}
\Qmin = \max_\spar \left\{ \Qpar \right\}.
\end{equation}
A straightforward approach to solving the above problem is to use the bisection algorithm~\cite{NocedalWright_NumericalOptimization}, where for each evaluated $\spar$ the value of $\Qpar$ is obtained by finding the smallest eigenvalue solution to \eqref{eq:eigenvalueDef}.

\section{Proof of a Tight Lower Bound}
\label{sec:Explanation}

The following discussion examines how the properties of the eigenspace associated with $\Ivec_\spar$ and $\Qpar$ always lead to equality in \eqref{eq:QpartsDef} for some value of $\spar$.

\subsection{One-dimensional Eigenspace}
\label{sec:Explanation:OneDimens}

\begin{figure}
\begin{center}
\includegraphics[width=9cm]{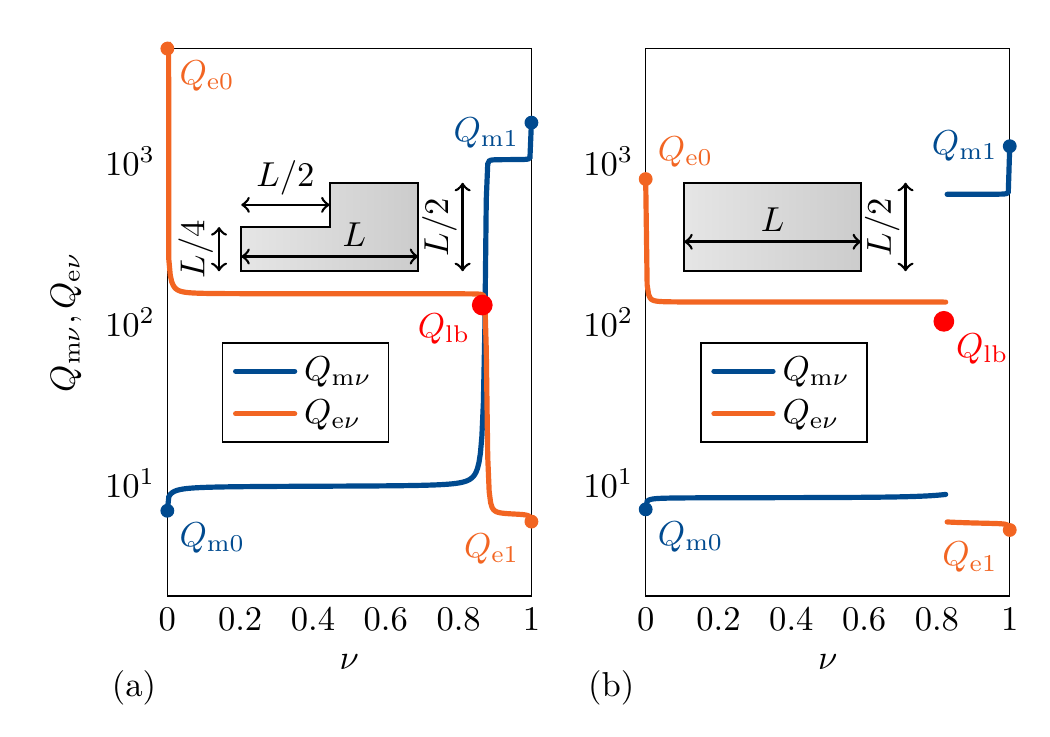}
\caption{The electric and magnetic modal quality factors $\QMpar$ and $\QEpar$ from \eqref{eq:QmQeDef} are depicted for two different shapes. The L-shape without any degeneracy is depicted on the left part. As all the curves are continuous, the minimal quality factor~$\Qmin$ can be found for such $\spar$ where \mbox{$\QMpar=\QEpar$}. The rectangular plate is depicted on the right part of the figure. The minimal quality factor~$\Qmin$ can be found in case the degenerate eigenspace is taken into account.}
\label{fig:ExplanF1}
\end{center}
\end{figure}

If the quality factor $\Qpar$ has one continuous associated eigenvector for all $\spar$ and at some $\nu_0$ this eigenvector is self resonant, \ie, \mbox{$Q_{m\nu_0}= Q_{e\nu_0}$}, then equality holds in \eqref{eq:QpartsDef} at \mbox{$\nu = \nu_0$}.  Figure~\ref{fig:ExplanF1}a depicts this self-resonant point for an L-shaped plate with largest dimension \mbox{$L = 0.1\lambda$}, where $\lambda$ denotes the wavelength. By \eqref{eq:QpartsDef}, $\Qmin$ is given by the value of $\QEpar$ and $\QMpar$ at their intersection, shown as a red dot in Fig.~\ref{fig:ExplanF1}a.

If the quality factor~$\Qpar$ has one continuous associated eigenvector for all $\spar$ but self-resonance is never achieved, (implying \mbox{$\QMpar > \QEpar$} for all $\spar$ or \mbox{$\QEpar > \QMpar$} for all $\spar$) then equality in \eqref{eq:QpartsDef} must hold at either \mbox{$\nu = 0$} or \mbox{$\nu = 1$}.  Hence, when $Q_\nu$ is always associated with a one-dimensional eigenspace, there exists a~$\spar$ which leads to equality in \eqref{eq:QpartsDef}.

\subsection{Instantaneous Degeneracies (Crossings)}
\label{sec:Explanation:InstantDegen}

In cases where $\Qpar$ has two or more degenerate eigenvectors at a discrete value $\nu_0$, the right hand side of (\ref{eq:QpartsDef}) remains single-valued while the left hand side of (\ref{eq:QpartsDef}) is not uniquely defined. This is because at the degeneracy $\Ivec_\spar$ can be any arbitrary vector within the multidimensional eigenspace associated with $\Qpar$. We can always choose a vector from this eigenspace which satisfies the self-resonant condition, provided the eigenvectors at $\nu_0-\delta\nu$ and \mbox{$\nu_0+\delta\nu$} have opposite ordering of $\QE$ and $\QM$.  Satisfying the self-resonance condition in this way implies equality in \eqref{eq:QpartsDef} at $\spar_0$.  In the specific case of a two-dimensional eigenspace with basis vectors $\Ivec_1$ and $\Ivec_2$, the self-resonant eigenvector $\Ivec_{\nu,\mathrm{sr}}$ is given by
\begin{equation}
\label{eq:DegTreat1}
\Ivec_{\spar,\mathrm{sr}} = \Ivec_1 + \mixModeAmp \mathrm{e}^{\J \mixModePhase}\Ivec_2
\end{equation}
where the real-valued coefficient $\chi$ is the solution to 
\begin{equation}
\label{eq:DegTreat2}
 \mixModeAmp^2 \Delta_{22} + 2  \mixModeAmp  \cos\left( \mixModePhase \right) \Delta_{12} + \Delta_{11} = 0
\end{equation}
and
\begin{equation}
\label{eq:DegTreat3}
\Delta_{mn} = \Ivec_m^\trans \left( \Xm - \Xe \right) \Ivec_n
\end{equation}
with the superscript $^\trans$ denoting the matrix transposition. Note that discrete degeneracies such as these occur only when a surface's geometric symmetries are precisely represented \cite{vonNeumannWigner_OnTheBehaviourOfEigenvaluesENG, SchabBernhard_GroupTheoryForCMA, Knorr_1973_TCM_symmetry}, in which case $\Delta_{12}$, by way of symmetry, is zero if $\Ivec_1$ and $\Ivec_2$ are chosen to be continuous with the eigenvectors at \mbox{$\nu_0-\delta\nu$} and \mbox{$\nu_0+\delta\nu$}.  From this we see that the phase associated with the combination in \eqref{eq:DegTreat1} is arbitrary.  

If the eigenvectors at \mbox{$\nu_0-\delta\nu$} and \mbox{$\nu_0+\delta\nu$} do not have opposite ordering of $Q_e$ and $Q_m$, then \eqref{eq:DegTreat2} has no real valued solution and the degeneracy can be otherwise ignored.  In this case, the rules for $Q_\nu$ with one-dimensional eigenspaces can be used to determine that there exists some $\spar$ where equality holds in \eqref{eq:QpartsDef}.

Figure~\ref{fig:ExplanF1}b shows $\QMpar$ and $\QEpar$ for a \mbox{$0.1 \lambda \times 0.05\lambda$} rectangular plate.  Due to the geometric symmetry of the plate, $\Qpar$ has a multidimensional eigenspace at \mbox{$\spar \approx 0.82$}, where $\QMpar$ and $\QEpar$ are discontinuous.  At this value of $\spar$, we can apply the preceding method of finding a self-resonant $\Ivec_\spar$ within the degenerate eigenspace. Using this self-resonant vector, the value of $\Qmin$ is given by \eqref{eq:QpartsDef}, shown as a red dot in Fig. \ref{fig:ExplanF1}b.

\subsection{Distributed Degeneracies}
\label{sec:Explanation:DistribDegen}

If $Q_\spar$ has a multi-dimensional eigenspace for all $\spar$, then $\Ivec_\spar$ should be selected as a vector which minimizes $\max \{Q_{m\spar},Q_{e\spar} \}$ within the eigenspace.  With the eigenvector defined in this way, the rules for $Q_\spar$ with one-dimensional eigenspaces can again be used to determine that equality must hold in \eqref{eq:QpartsDef} for some $\spar$.  This case is very rare in practice and will only occur when certain geometric symmetries are precisely represented, \eg, a square with rectangular mesh or a sphere with global basis functions.  The latter of these is discussed further in Section \ref{sec:Res:Ex3}.

\section{Generalization of the Minimization Problem}
\label{sec:Generalization}

The results of the previous sections are valid for arbitrary positive semi-definite matrices \mbox{$\Xm,\Xe,\M{R}$} and are hence directly generalized to several important antenna problems. For notational convenience, we rewrite the generalized eigenvalue problem~\eqref{eq:eigenvalueDef} as
\begin{equation}
\big( \left(1 - \spar \right) \gXm + \spar \gXe \big) \Ivec_\spar = \Upsilon_\spar \genRop \Ivec_\spar.
\label{eq:eigenvalueGenProb}
\end{equation}
In addition to the radiated power and stored energy, we consider quadratic forms enumerated in Table~\ref{Tab:GeneralizationExamples}. These quantities are combined with the stored energy~\eqref{eq:WmDef} and~\eqref{eq:WeDef} and radiated power~\eqref{eq:PrDef} to form eigenvalue problems for antennas with prescribed radiation patterns, maximization of $G/Q$, near-field constraints, and tradeoffs between stored energy and ohmic losses.
\begin{table}
\begin{center}
\begin{tabular}{|l|c|c|c|}
\hline 
\rule[1ex]{0pt}{1ex}
LHS of \eqref{eq:eigenvalueGenProb} & Notation & Bilinear form & Ref. \\ 
\hline 
\hline 
\rule[2ex]{0pt}{2ex}
Mode expansions & $P_\alpha$ & $\displaystyle\frac{1}{2\ZVAC} \ABS{\M{F}_{\alpha}\Ivec}^2$ & \eqref{eq:FFspherical} \\ 
	\hline 
\rule[2ex]{0pt}{2ex}    
Radiation intensity & $U(\rvh,\evh)$ & $\displaystyle\frac{1}{2\ZVAC}\ABS{\M{F}\Ivec}^2 = \frac{1}{2}\Ivec^\herm\M{U}\Ivec$ & \eqref{eq:FFfromRWG} \\ 
	\hline 
\rule[2ex]{0pt}{2ex}
Ohmic losses & $P_\sigma$ & $\displaystyle\frac{1}{2}\Ivec^\herm \M{R}_\sigma \Ivec$ & \eqref{eq:paretolosses} \\ 
	\hline 
\rule[2ex]{0pt}{2ex}
Field intensities & $|\evh\cdot\V{E}(\V{r})|^2$ & $|\M{N}_\mathrm{e}\Ivec|^2=\Ivec^\herm\M{N}_\mathrm{e}^\herm\M{N}_\mathrm{e}\Ivec$ & \cite{GustafssonTayliEhrenborgEtAl_AntennaCurrentOptimizationUsingMatlabAndCVX} \\ 
	\hline 
\end{tabular} 
\label{Tab:GeneralizationExamples}
\caption{Summary of additional quadratic forms which can be used to further constrain the radiated field, lost power or the near-field.}
\end{center}
\end{table}

The classical bounds on the quality factor~$Q$ are formulated for the electric dipole (TM), magnetic dipole (TE), and combined  cases~\cite{Chu_PhysicalLimitationsOfOmniDirectAntennas,VolakisChenFujimoto_SmallAntennas,Hansen_ESA_SuperDirAndSuperconducAntennas,GustafssonTayliCismasu_PhysicalBoundsOfAntennasBook}. The TM and TE cases are determined from an expansion of the radiated field in spherical modes and division of the radiated power~\eqref{eq:PrDef} in its TM and TE parts. Expansion of the radiated fields in spherical modes defines row matrices $\M{F}_{\alpha}$ with elements~\cite{Gustafsson_OptimalAntennaCurrentsForQsuperdirectivityAndRP}
\begin{equation}
F_{n, \sphi} = k \ZVAC \int\limits_{\srcRegion} \V{v}_{\sphi} \left(\V{r}'\right)\cdot\V{\psi}_n \left(\V{r}' \right) \D{S'},
\label{eq:FFspherical}
\end{equation}
where $\ZVAC$ is the free space impedance, \mbox{$k = 2\pi/\lambda$} is the wave-number, $\V{v}_{\sphi}$ denotes the real valued regular spherical vector waves \cite{Kristensson_ScatteringBook}, $\sphi$ is the multi-index of the mode and $\V{\psi}_n$ are the basis functions used in \eqref{eq:IvecDef}. The decomposition of TE and TM modes corresponds to a decomposition of the resistance matrix in its TE and TM parts 
\mbox{$\M{R}=\M{R}_\mathrm{TE}+\M{R}_\mathrm{TM}$}. The eigenvalue problem for minimizing quality factor~$Q$ for antennas radiating only TM fields is~\eqref{eq:eigenvalueGenProb} with 
\begin{equation}
\genRop = \M{R}_\mathrm{TM} = \frac{1}{2\ZVAC} \sum\limits_{\mathrm{TM~modes}} \M{F}_\alpha^\herm \M{F}_\alpha.
\label{eq:TMradPat}
\end{equation}

The radiation pattern and the radiated power can also be expressed in the partial radiation intensity $U(\rvh,\evh)$ for the direction $\rvh$ and polarization $\evh$. The partial radiation intensity $U(\rvh,\evh)$ is further rewritten using the far-field component \mbox{$\evh^{\ast}\cdot\V{F}(\rvh)=\M{F}(\rvh,\evh)\Ivec$}, where the complex conjugate is used for elliptic and circular polarized cases. The row matrix $\M{F}$ has the elements
\begin{equation}
F_n(\rvh,\evh) = -\frac{\J k \ZVAC}{4 \pi} \int\limits_\srcRegion \evh^\ast \cdot\V{\psi}_n\left(\V{r}'\right) \mathrm{e}^{\J k \rvh \cdot \V{r}'} \D{S'}
\label{eq:FFfromRWG}
\end{equation}
and \mbox{$U=\Ivec^{\herm}\M{U}\Ivec / 2$} with the matrix $\M{U}$
\begin{equation}
\M{U} \left(\rvh,\evh \right) = \frac{1}{\ZVAC}\M{F}^\herm \left(\rvh,\evh \right) \M{F}\left(\rvh,\evh\right).
\label{eq:Um}
\end{equation}
The minimum stored energy for an antenna having the radiation intensity weighted with $w_1$ and $w_2$ is given by~\eqref{eq:eigenvalueGenProb} with
\begin{equation}
\genRop = \frac{1}{2}\int\limits_{4\pi} \left( w_1(\rvh') \M{U}(\rvh',\evh_1) + w_2(\rvh') \M{U}(\rvh',\evh_2) \right) \D{S}'
\label{eq:Umw}
\end{equation}
where \mbox{$\evh_2=\rvh\times\evh_1^{\ast}$}. An alternative interpretation for the case with $w_1(\rvh)$ and $w_2(\rvh)$ as the characteristic functions for an angular region is as the minimum quality factor~$Q$ for antennas with negligible radiation outside the angular region. 

The particular choice of \mbox{$w_1(\rvh') = w_2(\rvh') = \delta(\rvh'-\rvh)$} in~\eqref{eq:Umw} gives an optimization problem for the quotient between quality factor~$Q$ and antenna gain $G$
\begin{equation}
G(\rvh) = \frac{4\pi U(\rvh)}{\Prad} = \frac{4\pi\Ivec^\herm\big(\M{U}(\rvh,\thetavh)+\M{U}(\rvh,\phivh)\big)\Ivec}{\Ivec^\herm\M{R}\Ivec}
\label{eq:Gain}
\end{equation}
The optimization problem for lower bounds on $Q/G$ or equivalently upper bounds on $G/Q$ is hence reformulated as minimization of  
\begin{equation}
\frac{Q}{G(\rvh)} = \frac{\max\{\Ivec^\herm\Xm\Ivec,\Ivec^\herm\Xe\Ivec\}}{4\pi\Ivec^{\herm}\big(\M{U}(\rvh,\thetavh)+\M{U}(\rvh,\phivh)\big)\Ivec}
\label{eq:QoG}
\end{equation}
where the spherical unit vectors $\thetavh$ and $\phivh$ are used to express the polarization. The generalized eigenvalue problem~\eqref{eq:eigenvalueGenProb} contains the rank two matrix
\begin{equation}
 \genRop = 4\pi\big(\M{U}(\rvh,\thetavh)+\M{U}(\rvh,\phivh)\big).
\end{equation}

It is also possible to use the dissipated power in the near field. Near fields are of interest for exposure, wireless power transfer, mutual coupling, and isolation \cite{GustafssonFridenColombi_AntennaCurrentOptimizationForLossyMediaAWPL}. Here, the matrix $\genRop$ is expressed as quadratic forms constructed by the near-field matrices $\M{N}_{\mathrm{e},n}$ and $\M{N}_{\mathrm{m},n}$ that are used to expressed the electric and magnetic near fields as \mbox{$\V{E}(\V{r}_n)=\M{N}_{\mathrm{e},n}\Ivec$} and \mbox{$\V{H}(\V{r}_n)=\M{N}_{\mathrm{m},n}\Ivec$}, respectively.

Ohmic losses can be included to investigate bounds on the radiation efficiency $\eta_\mathrm{rad}$ of antennas. For lossy antennas it desired to have both low losses and low stored energy for a fixed radiated power. This leads to a multi-objective optimization problem which can be analyzed using Pareto fronts~\cite{Gustafsson_ParetoOptimalityAPS} by minimization of
\begin{equation}
\frac{\parpar\max\{\Ivec^{\herm}\Xe\Ivec,\Ivec^{\herm}\Xm\Ivec\} + (1-\parpar)\Ivec^{\herm}\M{R}_{\sigma}\Ivec}{\Ivec^{\herm}\M{R}\Ivec}  
\label{eq:paretolosses}
\end{equation}  
for fixed value of the parameter $0\leq\parpar\leq 1$, where $\Ivec^{\herm}\M{R}_{\sigma}\Ivec$ is constructed to represent relevant dissipative losses in the system \cite{ChewTongHu_IntegralEquationMethodsForElectromagneticAndElasticWaves, JelinekCapek_OptimalCurrentsOnArbitrarilyShapedSurfaces}. This problem has the same form as~\eqref{eq:eigenvalueGenProb} with the substitutions $\genRop=\M{R}$ and
\begin{subequations}
\begin{equation}
  \gXe = \parpar\Xe + (1 - \parpar) \M{R}_{\sigma} 
\end{equation}
\begin{equation}
  \gXm = \parpar\Xm + (1 - \parpar) \M{R}_{\sigma}
\end{equation}
\end{subequations}
and is hence easily solved. The parameter $\parpar$ determines the weight between the stored energy and losses. Small and large values of $\parpar$ emphasizes low losses, expressed as \mbox{$\eta_\mathrm{rad}^{-1} - 1$}, and low $Q/\eta_\mathrm{rad}$, respectively. There are several other interesting multi-objective formulations, \eg{}, combining the $Q/G$ in~\eqref{eq:QoG} with $1/G$ in~\eqref{eq:Gain} as
\begin{equation}
 \frac{\parpar\max\{\Ivec^\herm\Xe\Ivec,\Ivec^\herm\Xm\Ivec\} + \left(1-\parpar\right)\Ivec^{\herm}\M{R}\Ivec}{4\pi\Ivec^{\herm}\big(\M{U}(\rvh,\thetavh) + \M{U}(\rvh,\phivh)\big)\Ivec}  
\end{equation}  
can be used to investigate the trade-off between quality factor~$Q$ and $G$. This is similar to the convex optimization formulation for minimum quality factor~$Q$ for superdirective antennas~\cite{Gustafsson_OptimalAntennaCurrentsForQsuperdirectivityAndRP}.

\section{Results}
\label{sec:Res}

In this section, the procedure to calculate the minimum quality factor~$\Qmin$ is demonstrated on a series of numerical examples of various complexity using bodies made of perfect electric conductor (PEC). Important consequences are discussed in Section~\ref{sec:Disc}.

First, a structure without geometric symmetry is treated in Section~\ref{sec:Res:Ex1}. It is shown that these cases are easily solvable, \cf{} Fig.~\ref{fig:ExplanF1}a. Second, a structure possessing geometric symmetries is analyzed in Section~\ref{sec:Res:Ex2} and the solution is found in the degenerate eigenspace, \cf{} Fig.~\ref{fig:ExplanF1}b. The spherical shell is considered in Section~\ref{sec:Res:Ex3} as the third example, being a representative of canonical bodies which are solvable analytically \cite{CapekJelinek_OptimalCompositionOfModalCurrentsQ}. The spherical shell is of great interest here as it possesses the highest degree of symmetry in $\mathbb{R}^3$. The last example in Section~\ref{sec:Res:Ex4} deals with a simple structure subject to sub-domain optimization.

All dimensions are normalized to $kL$ or, whenever proper, to $ka$, in which $L$ is a length of the structure (to be specified for each example individually) and $a$ is the radius of the smallest sphere completely surrounding the sources.

\subsection{L-shaped Plate}
\label{sec:Res:Ex1}

\begin{figure}
\begin{center}
\includegraphics[width=9cm]{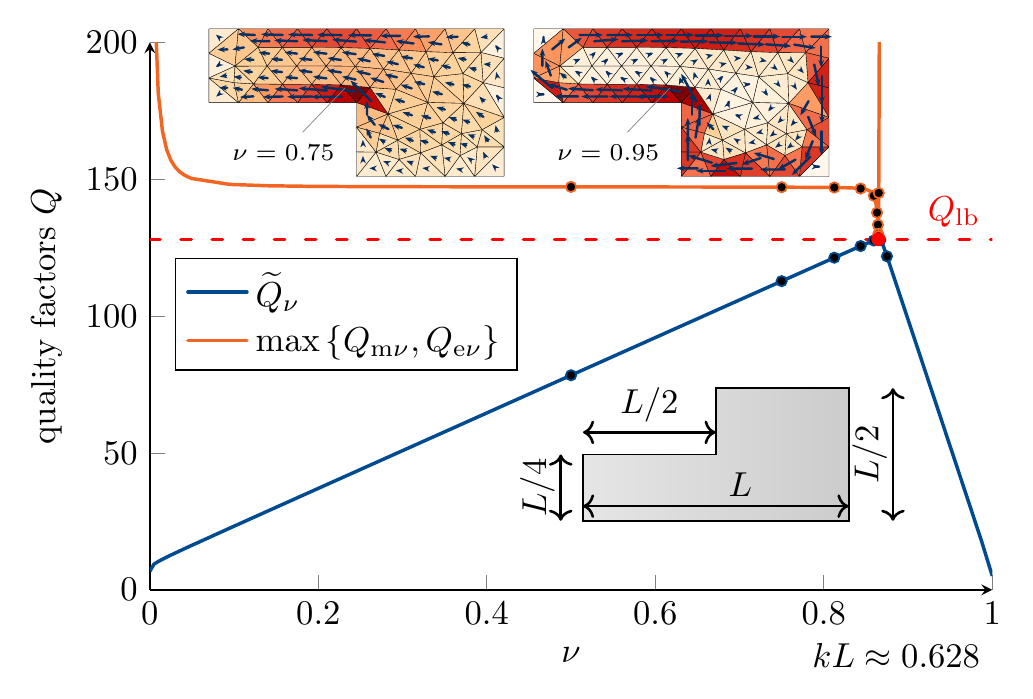}
\caption{The quality factors of the original and the dual problems, $\Qmin$ and $\Qdgmin$, for the L-shape, depicted as the inset on the right bottom, and electrical size \mbox{$kL \approx 0.628$} ($L=0.1\lambda$). The dot markers represent points at which the bisection algorithm determined to solve the problem \eqref{eq:eigenvalueDef}. Two insets at the top depict corresponding current density for \mbox{$\spar = \left\{0.75, 0.95\right\}$}.}
\label{fig:Ex1F1}
\end{center}
\end{figure}

A planar, L-shaped PEC structure 
(\,\begin{tikzpicture}[scale=0.50]
\draw [thick, left color=\cStructA, right color=\cStructB, line width=0.25pt] (0,0) -- (1,0) -- (1,0.5) -- (0.5,0.5) -- (0.5,0.25) -- (0,0.25) -- cycle;
\end{tikzpicture}\,) of length $L$, width $L/2$, cut of \mbox{$L/2 \times L/4$} and electrical size \mbox{$kL \approx 0.628$} is discretized into $792$ triangular elements and considered here as the first example, see the inset in Fig.~\ref{fig:Ex1F1}. 

\begin{figure}
\begin{center}
\includegraphics[width=9cm]{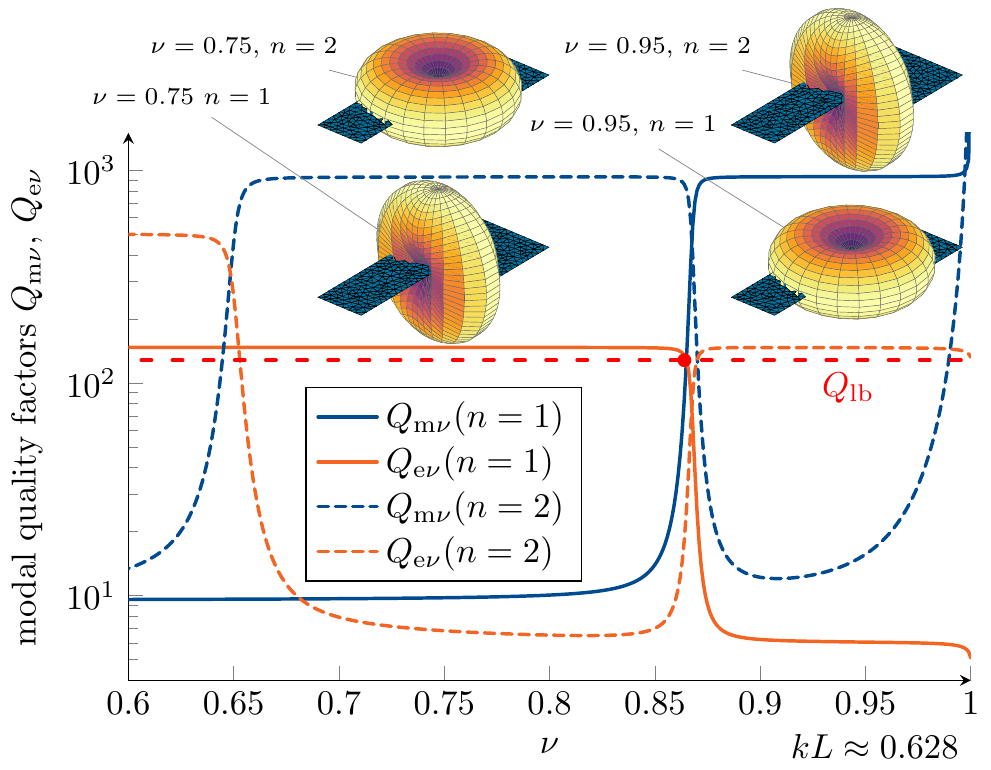}
\caption{Modal quality factors $\QMpar$ and $\QEpar$ for the L-shape from Fig.~\ref{fig:Ex1F1} evaluated at electrical size \mbox{$kL \approx 0.628$}. To put the behavior of $\Ivec_\spar$ into a broader context, the first two modes of (\ref{eq:eigenvalueDef}), denoted here by indices \mbox{$n \in \left\{ 1,2\right\}$}, are studied. Horizontal axis was limited to the relevant range of \mbox{$\nu \in \left[0.6,1\right]$}. The insets placed at the top depict corresponding radiation patterns for \mbox{$n = \left\{1, 2\right\}$} and \mbox{$\spar = \left\{0.75, 0.95\right\}$}. The density of mesh grid was scaled down in order to produce lucid illustrations.}
\label{fig:Ex1F2}
\end{center}
\end{figure}

Because the structure possesses no geometrical symmetries, the procedure described in Section~\ref{sec:Method} can be followed without special treatment for degenerate modes. In order to find the parameter $\spar$ at which \mbox{$\Qmin = \Qdgmin$}, the bisection algorithm~\cite{NocedalWright_NumericalOptimization} can be employed to solve\footnote{During each iteration we know the actual values of $\QMpar$ and $\QEpar$ and we know that these values are monotonically increasing/decreasing, see Fig.~\ref{fig:Ex1F2}.} the maximization problem in \eqref{eq:maxProblemDef}. The minimum quality factor~$\Qmin$ occurs at \mbox{$\spar\approx 0.865$}, see Fig.~\ref{fig:Ex1F1}. It reaches \mbox{$\Qmin = 128$} and yields \mbox{$\QMpar = \QEpar$}, see Fig.~\ref{fig:Ex1F2} as well. The radiation patterns of the current $\Ivec_\spar$ are shown in Fig.~\ref{fig:Ex1F2} for \mbox{$\spar = 0.75$} and \mbox{$\spar = 0.95$}, \ie, for values where \mbox{$\QMpar  < \QEpar$} and \mbox{$\QMpar > \QEpar$}, respectively. 

The behavior of the separate $\QMpar$ and $\QEpar$ terms is depicted in Fig.~\ref{fig:Ex1F2}. It can be seen that in the vicinity of the optimal solution the quality factors $\QMpar$ and $\QEpar$ change abruptly but still smoothly. The sharp minimum of quality factor~$\Qmin$ in Fig.~\ref{fig:Ex1F1} is direct consequence of this rapid change. Notice that because no symmetries are present on this structure, the same results can be obtained by, \eg, employing rectangular mesh elements. 

\subsection{Rectangular Plate}
\label{sec:Res:Ex2}

A structure with geometrical symmetries, a rectangular plate, is treated as the second example. This object remains unchanged under two axes of reflection and $180^\mathrm{o}$ rotation. Hence, it has a non-trivial symmetry group \cite{SchabBernhard_GroupTheoryForCMA} and, as such, a degenerate eigen-space may occur during the solution of \eqref{eq:maxProblemDef}.

The PEC plate (\,\begin{tikzpicture}[scale=0.50]
\draw [thick, left color=\cStructA, right color=\cStructB, line width=0.25pt]  (0,0) -- (1,0) -- (1,0.5) -- (0,0.5) -- cycle;
\end{tikzpicture}\,) has dimensions \mbox{$L\times L/2$} and electrical size \mbox{$kL \approx 0.628$}. It is divided into 1540 rectangular elements and calculated in a method-of-moments code tailored to preserve the geometric symmetry of the plate.

\begin{figure}
\begin{center}
\includegraphics[width=9cm]{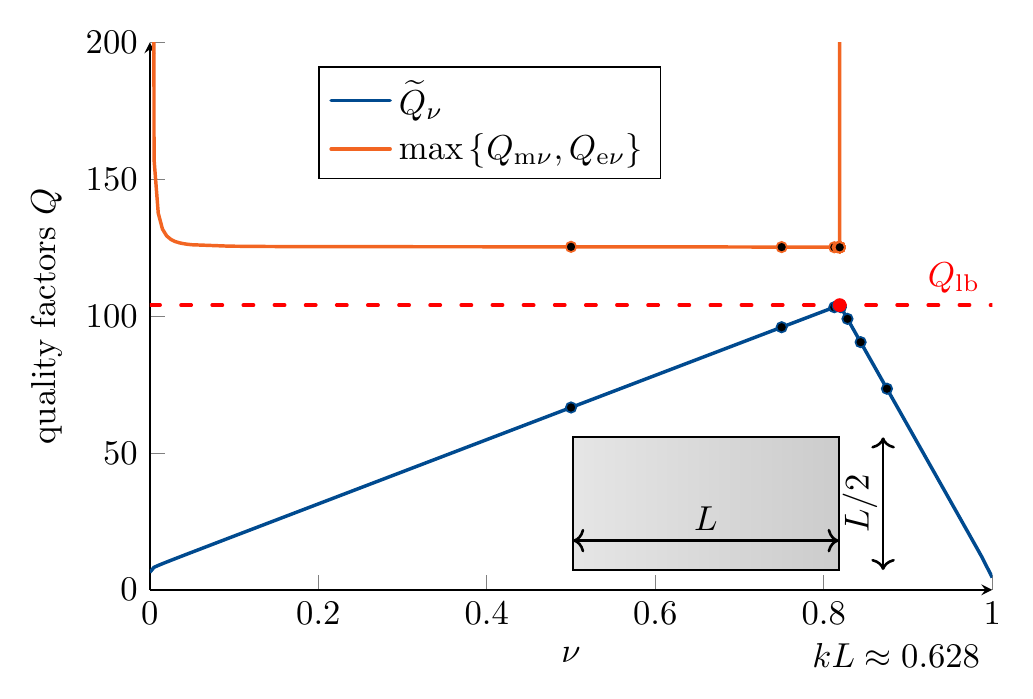}
\caption{The quality factors of the original and dual problems for the rectangular plate, depicted as the inset. The electrical size $kL$ and the meaning of dot markers are the same as in Fig.~\ref{fig:Ex1F1}.}
\label{fig:Ex2F1}
\end{center}
\end{figure}

When solving \eqref{eq:maxProblemDef}, a degenerate eigenspace is found at the maximum of $\Qpar$, see Fig. \ref{fig:Ex2F1}. As a consequence of the degeneracy, it is not immediately obvious that \mbox{$\Qmin = \Qdgmin$} at this point.  However, treating the degenerate eigenspace according to \eqref{eq:DegTreat1}--\eqref{eq:DegTreat3} one finds that this equality holds when the self-resonant eigenvector within the eigenspace is considered; leading to \mbox{$\Qmin = \Qdgmin = 103$}. The existence of the degenerate eigenspace introduces the additional degree of freedom for determination of the optimal current. This additional parameter, denoted in \eqref{eq:DegTreat1} as $\mixModePhase$, is the phase between degenerate eigenmodes and its importance is studied analytically in the next example. The appearance of the degeneracy in this example results from the strict enforcement of the plate's symmetry.  By perturbing the simulation through rounding errors or by using triangular basis functions \cite{RaoWiltonGlisson_ElectromagneticScatteringBySurfacesOfArbitraryShape, ChewTongHu_IntegralEquationMethodsForElectromagneticAndElasticWaves} which may not maintain symmetry, the degeneracy can be effectively eliminated.

\subsection{Spherical Shell}
\label{sec:Res:Ex3}

The PEC spherical shell of radius $a$ is examined in this section. Spherical harmonics \cite{Stratton_ElectromagneticTheory} are used as basis functions for which the impedance matrix and the $\Xm$, $\Xe$, and $\M{R}$ matrices are diagonal and can be evaluated analytically~\cite{Gustaffson_StoredElectromagneticEnergy_PIER}.

\begin{figure}
\begin{center}
\includegraphics[width=9cm]{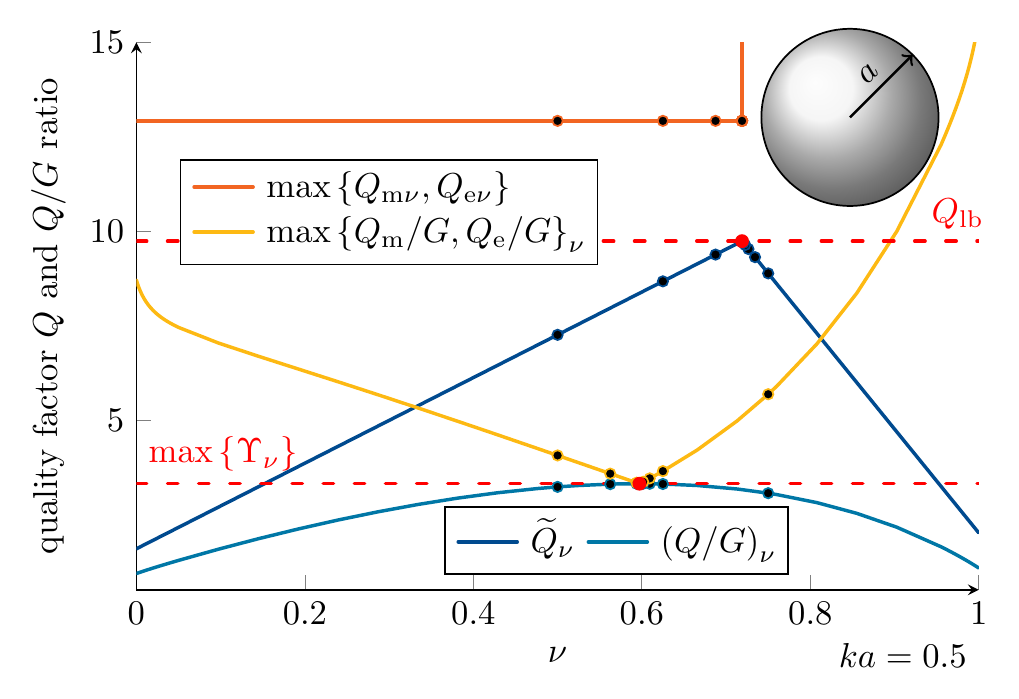}
\caption{The quality factors~$Q$ and $Q/G$ ratios for the spherical shell, depicted as the inset. Both solutions of the original and the dual problem are depicted. The electrical size is \mbox{$ka = 0.5$}. The dual problems are represented by blue lines, the primal problems are represented by orange lines. Similarly as in Fig.~\ref{fig:Ex1F1} and Fig.~\ref{fig:Ex2F1}, the dot markers explicitly depict values of $\spar$ at which the bisection algorithm determine to evaluate \eqref{eq:eigenvalueDef}.}
\label{fig:Ex3F1}
\end{center}
\end{figure}

The maximization procedure~\eqref{eq:maxProblemDef} is recapitulated in Fig.~\ref{fig:Ex3F1} for the particular choice of \mbox{$ka = 0.5$}. Since the eigenspace~\eqref{eq:eigenvalueDef} of the spherical shell is degenerated, a gap of width \mbox{$\Qmin - \Qdgmin$} occurs. It can be, however, effectively removed by using procedure from Section~\ref{sec:Explanation}. Interestingly, $Q_{\mathrm{m}\spar}$ and $Q_{\mathrm{e}\spar}$ are constant with respect to $\spar$, which means that $Q_\spar$ is a linear function of $\spar$. This is verified in Fig.~\ref{fig:Ex3F1} and simplifies locating the minimum quality factor \mbox{$\Qmin \approx 9.73$}, \cf{} \cite{CapekJelinek_OptimalCompositionOfModalCurrentsQ, Thal_RadiationQandGainOfTMandTEsourcesInPhaseDelayedRotatedConfigurations}.

The second case depicted in Fig.~\ref{fig:Ex3F1}, deals with maximization of $G/Q$, here reformulated as minimization of $Q/G$. There is no duality gap for this case, in agreement with the partial $G/Q$ case~\cite{GustafssonTayliEhrenborgEtAl_AntennaCurrentOptimizationUsingMatlabAndCVX} and the procedure gives \mbox{$Q/G \approx 3.31$} (\mbox{$G/Q \approx 0.302$}).

\begin{figure}
\begin{center}
\includegraphics[width=9cm]{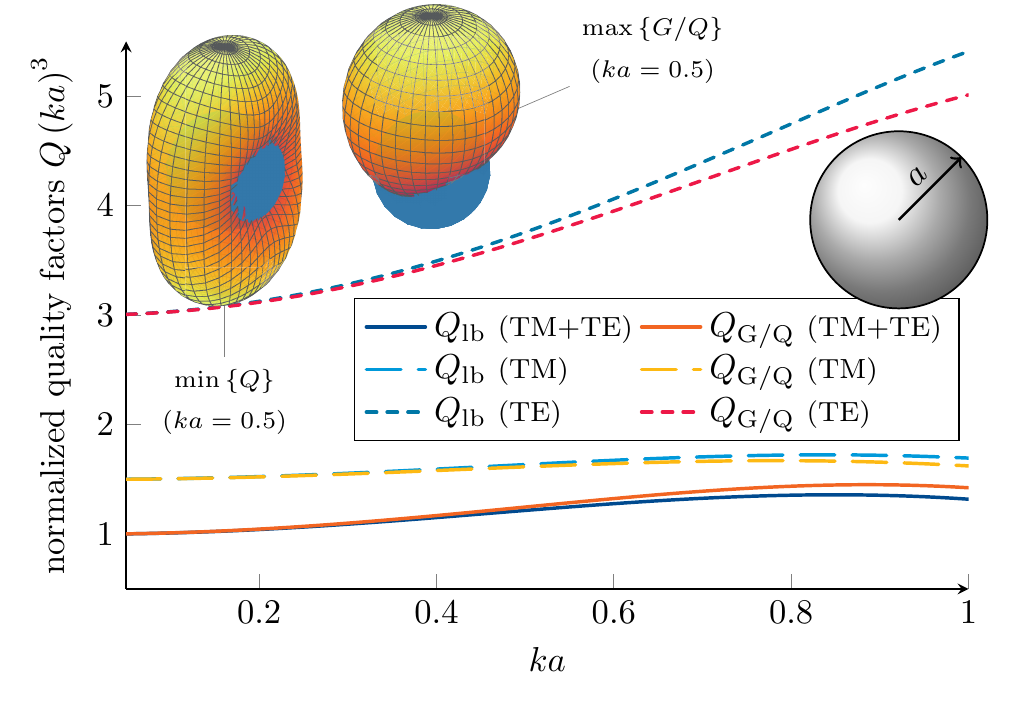}
\caption{Minimal quality factors~$Q$ for the spherical shell of radius $a$ studied in dependence on electrical size $ka$. The blue curve depict minimal quality factor~$\Qmin$, the orange curve depict quality factor~$Q_\mathrm{Q/G}$, corresponding to the minimal value of~$Q/G$. All curves are also categorized with respect to different radiation constrains: TM+TE stands for combination of electric and magnetic radiation modes, TM stand for electric (dipole-like) and TE for magnetic (loop-like) modes. All quality factors are normalized to $\left(ka \right)^3$ in order to flatten the curves.}
\label{fig:Ex3F2}
\end{center}
\end{figure}

The minimal quality factors $\Qmin$ and $Q_\mathrm{G/Q}$ are depicted in Fig.~\ref{fig:Ex3F2} as functions of the sphere's electrical size, $ka$. The quality factor~$Q_\mathrm{G/Q}$ is calculated from \eqref{eq:Qdef} with the current minimizing the ratio $G/Q$. The quality factors are depicted for the cases of arbitrary radiated fields (TM+TE), radiated electric modes (TM), and radiated magnetic modes (TE) calculated using~\eqref{eq:TMradPat}. The computed quality factors agree with the results in~\cite{CapekJelinek_OptimalCompositionOfModalCurrentsQ}. We observe that \mbox{$Q_\mathrm{G/Q}\approx \Qmin$} for $ka\ll 1$ and the slight increase of $Q_\mathrm{G/Q}$ over $\Qmin$ for larger $ka$ is due to excitation of higher order spherical modes, \ie, \mbox{$Q_\mathrm{G/Q}=\Qmin$} if the radiated fields are restricted to the dipole modes. Moreover, as the phase between modes can be chosen arbitrarily for the degenerate eigenspace~\eqref{eq:DegTreat1} the $G/Q$~ratio can be increased by proper selection of TM-TE mode doublet with optimal phase shift.

The spherical shell is classical example of a symmetrical body possessing distributed degeneracies. As explained in Section~\ref{sec:Explanation:DistribDegen}, this gives us additional freedom selecting the preferable optimal currents. For non-symmetrical body there is always unique current yielding minimum quality factor~$Q$ and having typically elliptical polarization. However, considering body with distributed degeneracies, we can freely select preferable polarization.

\subsection{Current Optimization in Sub-domain Region}
\label{sec:Res:Ex4}

The last example considers sub-domain antenna optimization~\cite{Gustafsson_OptimalAntennaCurrentsForQsuperdirectivityAndRP}, where the region $\srcRegion$ is divided into a controllable antenna region~$\srcRegion_\mathrm{A}$ and an uncontrollable ground region $\srcRegion_\mathrm{G}$.  Currents in $\srcRegion_\mathrm{A}$ (denoted $\Ivec_\mathrm{A}$) may be controlled via excitation voltages, whereas currents $\srcRegion_\mathrm{G}$ (denoted $\Ivec_\mathrm{G}$) are restricted to those induced by currents in $\srcRegion_\mathrm{A}$. We have \mbox{$\Ivec = \Ivec_\mathrm{A} \cup \Ivec_\mathrm{G}$} and \mbox{$\srcRegion = \srcRegion_\mathrm{A} \cup \srcRegion_\mathrm{G}$}, and consequently the discretized EFIE can be written in block matrix form as
\begin{equation}
\label{eq:SubdomOptim1}
\begin{array}{l}
\M{Z}_\mathrm{AA} \Ivec_\mathrm{A} + \M{Z}_\mathrm{AG} \Ivec_\mathrm{G} = \M{V}, \\
\M{Z}_\mathrm{GA} \Ivec_\mathrm{A} + \M{Z}_\mathrm{GG} \Ivec_\mathrm{G} = \M{0},
\end{array}
\end{equation}
leading to
\begin{equation}
\label{eq:SubdomOptim2}
\Ivec_\mathrm{G} = -\M{Z}_\mathrm{GG}^{-1} \M{Z}_\mathrm{GA} \Ivec_\mathrm{A} = \M{T} \Ivec_\mathrm{A}.
\end{equation}
Leveraging the division of controllable and uncontrollable currents and the relation in \eqref{eq:SubdomOptim2}, any quadratic form of the total current can be compressed into a quadratic form of the controllable current, \ie,
\begin{equation}
\label{eq:SubdomOptim3}
\Ivec^\herm \M{A} \Ivec = \Ivec_\mathrm{A}^\herm \mathcal{C} \left( \M{A} \right) \Ivec_\mathrm{A},
\end{equation}
where
\begin{equation}
\label{eq:SubdomOptim4}
\mathcal{C} \left( \M{A} \right) = \M{A}_\mathrm{AA} + \M{A}_\mathrm{AG} \M{T} + \M{T}^\herm \M{A}_\mathrm{GA} + \M{T}^\herm \M{A}_\mathrm{GG} \M{T}
\end{equation}
for any matrix $\M{A}$ operating on the current vector $\M{I}$.

\begin{figure}
\begin{center}
\includegraphics[width=9cm]{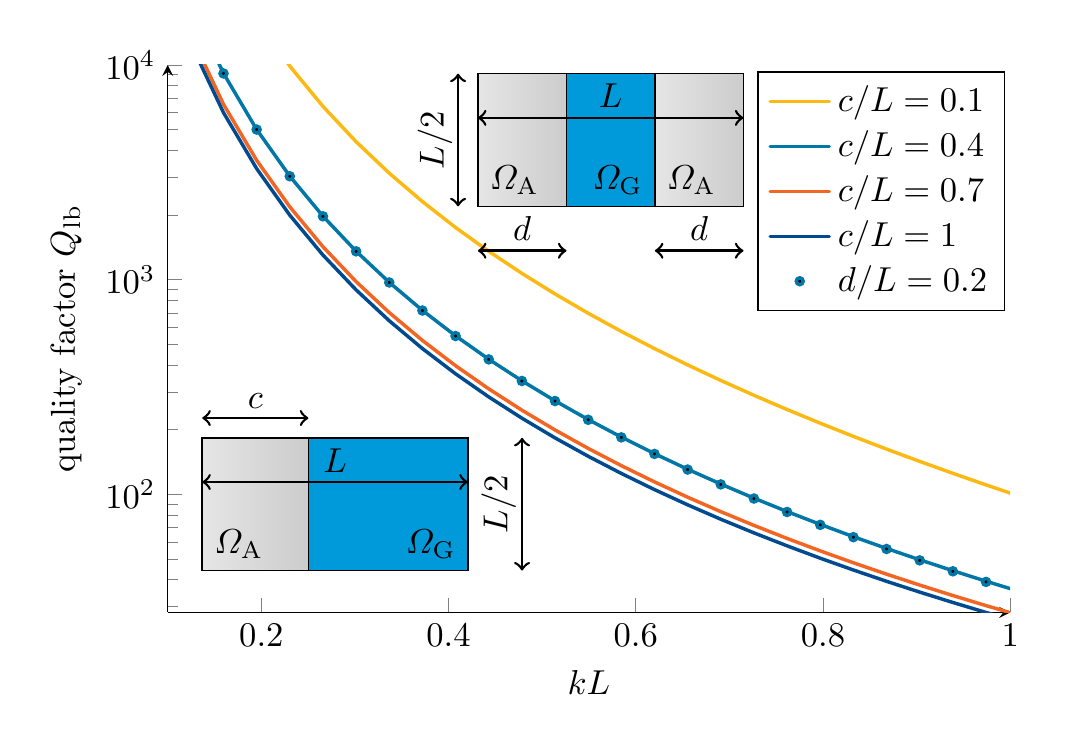}
\caption{The minimal quality factor~$\Qmin$ in dependence on electrical size $kL$ is depicted for several sub-domain optimization scenarios. The regions with controllable currents are grayed and denoted as $\srcRegion_\mathrm{A}$.}
\label{fig:Ex4F1}
\end{center}
\end{figure}

To investigate sub-domain optimization thoroughly, several different controllable regions are considered, see Fig.~\ref{fig:Ex4F1} and Table~\ref{Table1SubDomOpt}. In all cases, the mesh refinement of the controllable region $\srcRegion_\mathrm{A}$ is improved to provide enough degrees of freedom for optimal current. The RWG basis functions coinciding with the boundary between $\srcRegion_\mathrm{A}$ and $\srcRegion_\mathrm{G}$ were considered as controllable.

The rectangular plate of the same dimensions as in Section~\ref{sec:Res:Ex2} is treated in Fig.~\ref{fig:Ex4F1} and the dependence on size of controllable region and electrical size is studied. It can be seen that the dependence on the electrical size embodies the same course for different areas of region $\srcRegion_\mathrm{A}$. Practically the same quality factors~$\Qmin$ for curves \mbox{$c/L=0.1$} and \mbox{$d/L=0.2$} in Fig.~\ref{fig:Ex4F1} suggest that there is no significant difference between various sub-domain divisions while the constant ratio between areas $\srcRegion_\mathrm{A}$ and $\srcRegion_\mathrm{G}$ is preserved. It is, however, demonstrated in Table~\ref{Table1SubDomOpt} that this observation is not completely true -- interestingly, if the controllable region $\srcRegion_\mathrm{A}$ is well-chosen, the loss of control over 50\% of the total area led to only 30\% increase of minimal quality factor~$\Qmin$. Of particular interest is the third candidate in Table~\ref{Table1SubDomOpt}, for which the controllable region covers the outer loop of the rectangle. This region seems to be the optimal choice as we know that it resembles the loop-like optimal current, \cf{}~Fig.~\ref{fig:Ex2F1}. However, the value of the quality factor~$\Qmin$ is quite poor due to capacitive coupling between controllable and induced current. This is important observation, especially in the context of small antenna design, since often the parasitic ground plane is prescribed and the controllable region for antenna design is limited. Results presented in this work imply that there exists both an optimal shape and optimal placement of the controllable area.

\begin{table}
\begin{center}
\definecolor{colG}{rgb}{0,0.6,0.8549}
\def\rL{2}   
\def\rW{\rL/2} 
\begin{tabular}{|c||c|c|c|}
	\hline 
\rule{0pt}{2ex}    
optimized structure	& $\nu$ & $Q_\mathrm{lb}$ & to see corresponding $\Ivec_\spar$ \\ 
	\hline 	
	\hline 
\rule{0pt}{6ex}
\noindent\parbox[c]{\rL cm}{
\begin{tikzpicture}
\def\rD{0.5} 
\coordinate (A1) at ({-\rL/2},{-\rW/2});
\coordinate (B) at ({-\rL/2+\rD*\rL},{-\rW/2});
\coordinate (C) at ({-\rL/2+\rD*\rL},{\rW/2});
\coordinate (A2) at ({-\rL/2},{\rW/2});
\coordinate (G1) at ({\rL/2},{-\rW/2});
\coordinate (G2) at ({\rL/2},{\rW/2});

\draw [left color=black!10, right color=black!20,, line width=0.5pt] (A1) -- (B) -- (C) -- (A2) -- cycle;	
\draw [fill=colG, line width=0.5pt] (B) -- (G1) -- (G2) -- (C) -- cycle;	
\node[above right] at (A1) {$\srcRegion_\mathrm{A}$};
\node[above right] at (B) {$\srcRegion_\mathrm{G}$};
\end{tikzpicture}}
\rule[-5ex]{0pt}{0ex}
& $0.858$ & $131$ & Fig.~\ref{fig:Ex4F2}, top left \\ 
	\hline 
\rule{0pt}{6ex}
\noindent\parbox[c]{\rL cm}{
\begin{tikzpicture}
\def\rD{0.5} 
\coordinate (A1) at ({-\rL/2},{-\rW/2});
\coordinate (A2) at ({-\rL/2},{\rW/2});
\coordinate (G1) at ({-\rD},{-\rW/2});
\coordinate (G2) at ({-\rD},{\rW/2});
\coordinate (G3) at ({\rD},{-\rW/2});
\coordinate (G4) at ({\rD},{\rW/2});
\coordinate (A3) at ({\rL/2},{-\rW/2});
\coordinate (A4) at ({\rL/2},{\rW/2});

\draw [left color=black!10, right color=black!20,, line width=0.5pt] (A1) -- (G1) -- (G2) -- (A2) -- cycle;	
\draw [fill=colG, line width=0.5pt] (G1) -- (G3) -- (G4) -- (G2) -- cycle;	
\draw [left color=black!10, right color=black!20,, line width=0.5pt] (G3) -- (A3) -- (A4) -- (G4) -- cycle;
\end{tikzpicture}}
\rule[-5ex]{0pt}{0ex}
& $0.860$ & $130$ & Fig.~\ref{fig:Ex4F2}, bottom left \\ 
	\hline 
\rule{0pt}{6ex}
\noindent\parbox[c]{\rL cm}{
\begin{tikzpicture}
\def\rD{sqrt(2)/2} 
\coordinate (A1) at ({-\rL/2},{-\rW/2});
\coordinate (A2) at ({\rL/2},{-\rW/2});
\coordinate (A3) at ({\rL/2},{\rW/2});
\coordinate (A4) at ({-\rL/2},{\rW/2});

\coordinate (G1) at ({-\rD*\rL/2},{-\rD*\rW/2});
\coordinate (G2) at ({\rD*\rL/2},{-\rD*\rW/2});
\coordinate (G3) at ({\rD*\rL/2},{\rD*\rW/2});
\coordinate (G4) at ({-\rD*\rL/2},{\rD*\rW/2});

\draw [left color=black!10, right color=black!20, line width=0.5pt] (A1) -- (A2) -- (A3) -- (A4) -- cycle;	
\draw [fill=colG, line width=0.5pt] (G1) -- (G2) -- (G3) -- (G4) -- cycle;
\end{tikzpicture}}
\rule[-5ex]{0pt}{0ex}
 & $0.807$ & $162$ & Fig.~\ref{fig:Ex4F2}, top right \\ 
	\hline 
\rule{0pt}{6ex}
\noindent\parbox[c]{\rL cm}{
\begin{tikzpicture}
\def\rD{sqrt(2)/2} 
\coordinate (A1) at ({-\rL/2},{-\rW/2});
\coordinate (A2) at ({\rL/2},{-\rW/2});
\coordinate (A3) at ({\rL/2},{\rW/2});
\coordinate (A4) at ({-\rL/2},{\rW/2});

\coordinate (G1) at ({-\rD*\rL/2},{-\rD*\rW/2});
\coordinate (G2) at ({\rD*\rL/2},{-\rD*\rW/2});
\coordinate (G3) at ({\rD*\rL/2},{\rD*\rW/2});
\coordinate (G4) at ({-\rD*\rL/2},{\rD*\rW/2});

\draw [fill=colG, line width=0.5pt] (A1) -- (A2) -- (A3) -- (A4) -- cycle;	
\draw [left color=black!10, right color=black!20, line width=0.5pt] (G1) -- (G2) -- (G3) -- (G4) -- cycle;
\end{tikzpicture}}
\rule[-5ex]{0pt}{0ex}
& $0.594$ & $1400$ & Fig.~\ref{fig:Ex4F2}, bottom right \\
	\hline 
	\hline     
\rule{0pt}{6ex}
\noindent\parbox[c]{\rL cm}{
\begin{tikzpicture}
\def\rD{sqrt(2)/2} 
\coordinate (A1) at ({-\rL/2},{-\rW/2});
\coordinate (A2) at ({\rL/2},{-\rW/2});
\coordinate (A3) at ({\rL/2},{\rW/2});
\coordinate (A4) at ({-\rL/2},{\rW/2});

\draw [left color=black!10, right color=black!20, line width=0.5pt] (A1) -- (A2) -- (A3) -- (A4) -- cycle;
\node[above right] at (A1) {$\srcRegion_\mathrm{A}$};
\end{tikzpicture}}
\rule[-5ex]{0pt}{0ex}
 & $0.818$ & $103$ & -- \\ 
	\hline   
\end{tabular}
 \caption{Comparison of the minimal quality factors~$\Qmin$ for the rectangular area of dimension \mbox{$L \times L/2$} and electrical size \mbox{$kL \approx 0.628$}, divided into controllable area $\srcRegion_\mathrm{A}$ (gray) and area with induced currents only $\srcRegion_\mathrm{G}$ (blue). All samples were divided so that the half of the total area is controllable. The original rectangle with all currents controllable is added for completeness.}
\label{Table1SubDomOpt}
\end{center}
\end{table}

\begin{figure}
\begin{center}
\includegraphics[width=6.5cm]{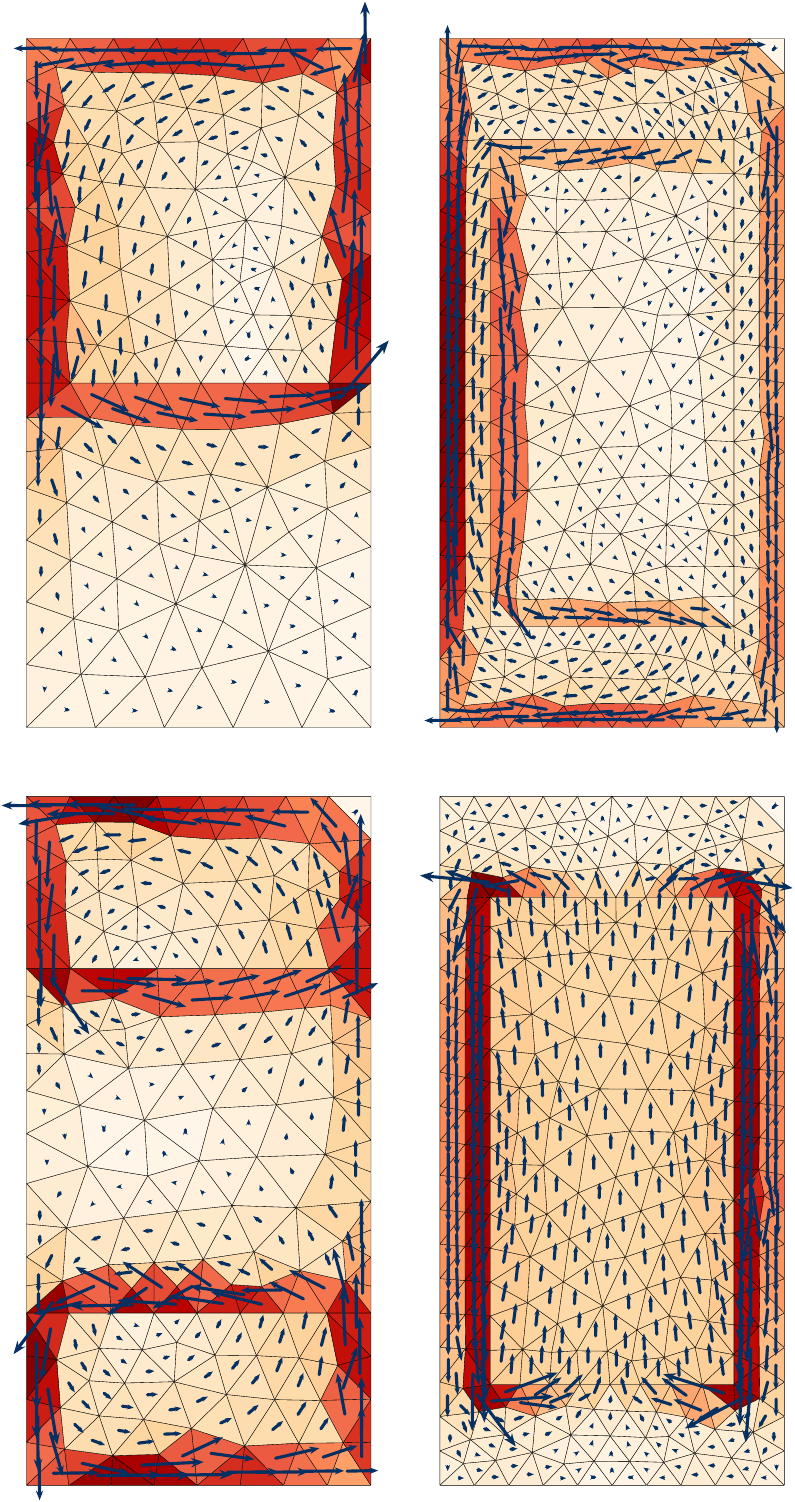}
\caption{Optimal currents for structures depicted in Table~\ref{Table1SubDomOpt} for the same electrical size (\mbox{$kL \approx 0.628$}). The optimal currents are real-valued. The arrows in the quiver plots have normalized length and they depict the direction of the current. The color maps depict the normalized absolute value of the current density at each triangle. The density of mesh grid was scaled down in order to produce lucid illustrations.}
\label{fig:Ex4F2}
\end{center}
\end{figure}

\section{Discussion}
\label{sec:Disc}

It has been proved in Section~\ref{sec:Explanation} and verified on practical examples in Section~\ref{sec:Res} that solving the maximization problem in~\eqref{eq:maxProblemDef} always minimizes quality factor~$Q$.  In Section~\ref{sec:Generalization}, other problems such as minimization of the $Q/G$ ratio, were discussed. Several points not yet discussed regarding the implementation and properties of this method are briefly treated in the remaining part of the paper.

\subsection{Requirements}
\label{sec:Disc:P1}

The impedance matrix $\M{Z}$ must be symmetric, \ie, \mbox{$\M{Z} = \left(\M{Z} + \M{Z}^\trans\right)/2$}. This is always true if Galerkin testing is adopted and properly implemented~\cite{ChewTongHu_IntegralEquationMethodsForElectromagneticAndElasticWaves}. Fortunately, this is a common choice when formulating the method of moments. In addition, it is required that the real part of the impedance matrix is positive semi-definite, \ie, \mbox{$\M{R} \succeq 0$}. This last requirement is grounded in the physical argument that any current distribution must radiate non-negative power.  Due to numerical errors, many method of moments codes produces a matrix $\M{R}$ with small negative eigenvalues.  These negative eigenvectors should be removed to enforce \mbox{$\M{R} \succeq 0$}, \cite{GustafssonTayliEhrenborgEtAl_AntennaCurrentOptimizationUsingMatlabAndCVX}.

From all presented results, we observe that the optimal current which minimizes quality factor~$Q$ tends to equalize magnetic and electric energies, \ie, \mbox{$Q_{e\spar} = Q_{m\spar}$}. Whether explicitly mixed at a discrete degeneracy or implicitly occurring at a smooth maximum when solving~\eqref{eq:maxProblemDef}, we can interpret the balance of magnetic and electric energies as blending of predominately electric (dipole-like) and magnetic (loop-like) type currents. It is important to note that the choice of discretization can directly impact the ability to represent loop-type currents, \eg, a straight wire dipole represented with one-dimensional basis functions (no circumferential flow) lacks the ability to support divergence-free loop currents.

\subsection{Properties}
\label{sec:Disc:P2}

As compared to other methods, it can be somewhat surprising that a purely real valued current is sufficient to deliver the minimum quality factor~$Q$, \cf~\cite{JelinekCapek_OptimalCurrentsOnArbitrarilyShapedSurfaces}. This is, however, well-aligned with recent attempts utilizing, \eg, characteristic modes,~\cite{CapekJelinek_OptimalCompositionOfModalCurrentsQ}. With respect to those modal approaches, this method can be understood in the same way -- as a method combining modes. However, because typically only one or two modes are needed, we can conclude that the solution here is represented in its natural basis.

The minimization procedure can be applied for any problem with the convex combination of stored energy operators on the LHS and a positive semi-definite matrix on the RHS of~\cite{GustafssonTayliEhrenborgEtAl_AntennaCurrentOptimizationUsingMatlabAndCVX}. Furthermore, and contrary to the previous attempts, the weighted matrix on the RHS is not restricted to being a rank-one matrix, therefore the class of antenna problems solvable by convex optimization is greatly enlarged. Notice, however, that the stored energy operators $\Xm$ and $\Xe$ in~\eqref{eq:WmDef} and~\eqref{eq:WeDef} fail to deliver the true stored energy for higher frequencies~\cite{GustafssonCismasuJonsson_PhysicalBoundsAndOptimalCurrentsOnAntennas_TAP}, \ie, \mbox{$ka > 1$}. In the context of this work, this fact is only a minor technicality and does not restrict the generality or applicability of the proposed method.

Finally, notice direct portability of the method to magnetic current or combination of electric and magnetic currents.

\section{Conclusion}
\label{sec:Concl}

The long-persistent problem of quality factor~$Q$ minimization for arbitrarily shaped body has been reformulated in convex form. Consequently, it has been solved using straightforward and computationally inexpensive procedure. As long as the required algebraic properties are guaranteed, the same procedure is directly applicable to other problems as well. Namely, the $Q/G$~ratio can easily be maximized (rewritten as minimization of $Q/G$), problems involving various near- or far-field restrictions can be solved, ohmic losses can be taken into account and the presence of regions with uncontrollable currents are now solvable. The enumerated problems are solved by the same formalism, which makes the potential implementation appealing for all scientists and other workers in the field.

One of the byproducts of the method is the link established between several diverse techniques which have been merged and further generalized. An important outcome of this work is the understanding that the current optimal with respect to quality factor~$Q$ and $G/Q$ have typically a uniquely defined inter-modal phase. For the special case of symmetric objects, this phase may be arbitrary for some quantities being minimized.

The presented approach also opens new questions to be addressed. For example, it seems clear that, when having control of currents over only part of a structure, there is an optimal shape of the controllable area to have the greatest impact on the antenna's performance. Another example is the open problem of controlling interactions within multi-antenna systems and the existence of optimal currents for these scenarios.

\section*{Acknowledgment}
The authors are grateful to Doruk Tayli (Lund University, Sweden) for a couple of numerical comparisons.
\ifCLASSOPTIONcaptionsoff
  \newpage
\fi

\bibliographystyle{IEEEtran}
\bibliography{references_LIST_UpToDate}

\begin{biography}[{\includegraphics[width=1in,height=1.25in,clip,keepaspectratio]{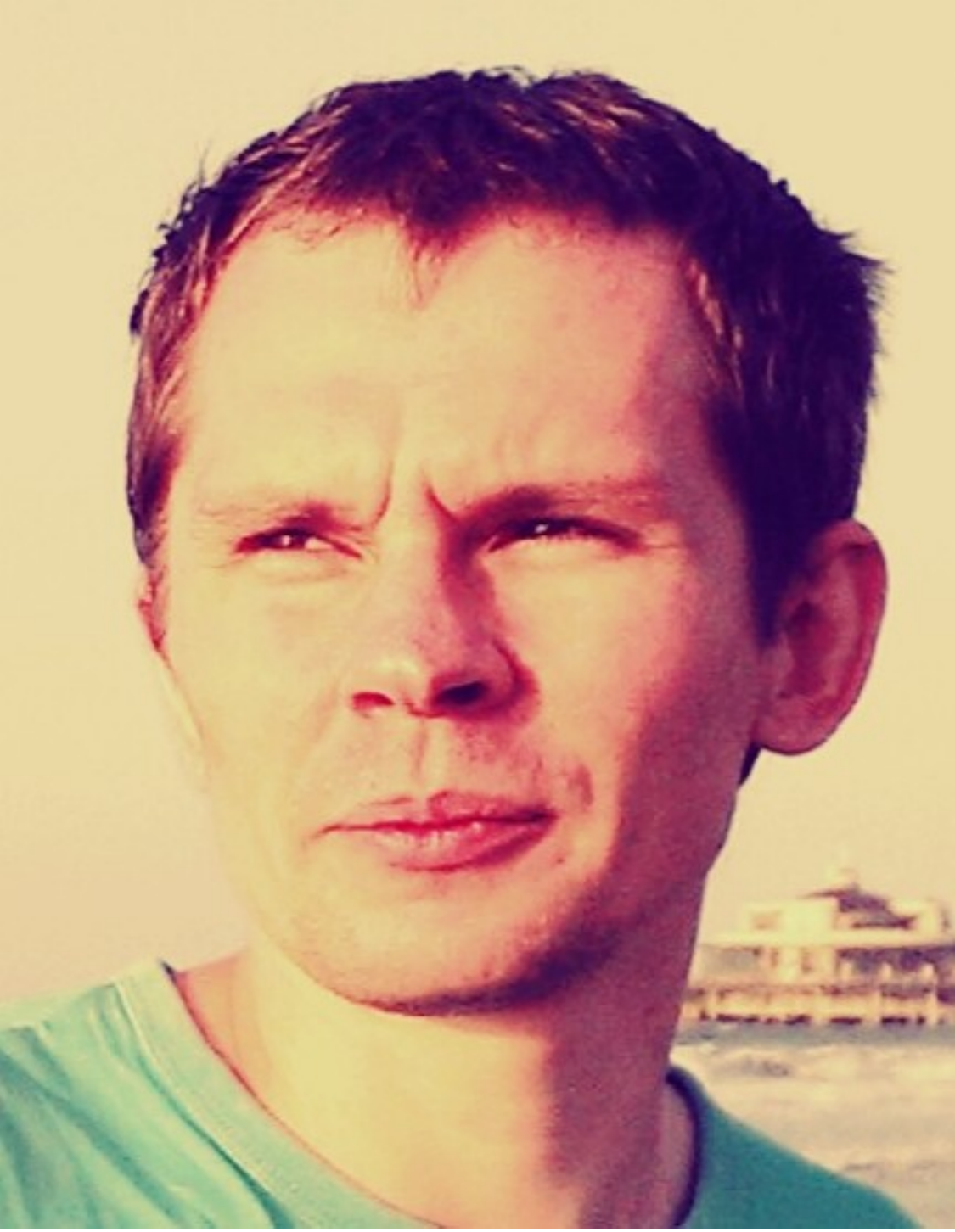}}]{Miloslav Capek}
(S'09, M'14) received his M.Sc. degree in Electrical Engineering from the Czech Technical University, Czech Republic, in 2009, and his Ph.D. degree from the same University, in 2014. Currently, he is a researcher with the Department of Electromagnetic Field, CTU-FEE.
	
He leads the development of the AToM (Antenna Toolbox for Matlab) package. His research interests are in the area of electromagnetic theory, electrically small antennas, numerical techniques, fractal geometry and optimization. He authored or co-authored over 50 journal and conference papers.

Dr. Capek is member of Radioengineering Society, regional delegate of EurAAP, and Associate Editor of Radioengineering.
\end{biography}

\begin{biography}[{\includegraphics[width=1in,height=1.25in,clip,keepaspectratio]{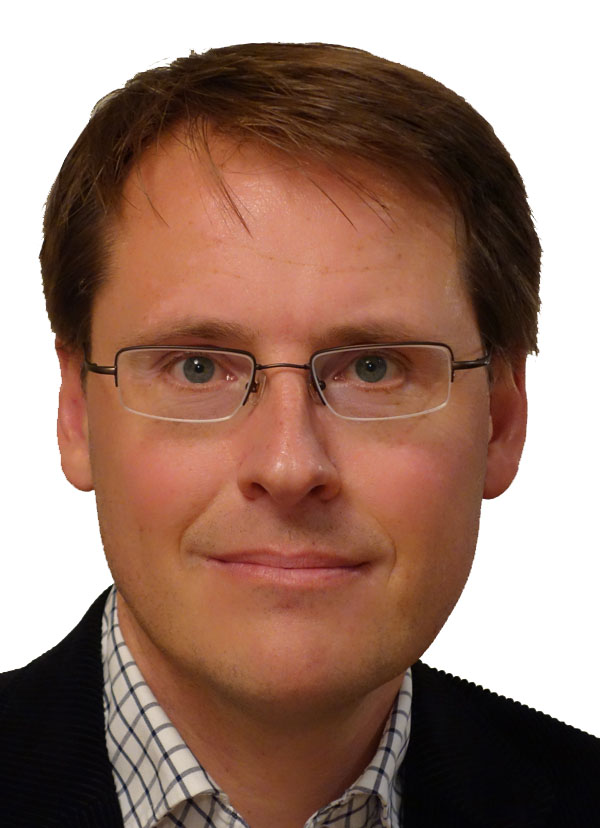}}]{Mats Gustafsson}
received the M.Sc. degree in Engineering Physics 1994, the Ph.D. degree in Electromagnetic
Theory 2000, was appointed Docent 2005, and Professor of Electromagnetic Theory 2011, all
from Lund University, Sweden. 

He co-founded the company Phase holographic imaging AB in 2004. His research interests are in scattering and antenna theory and inverse scattering and imaging. He has written over 80 peer reviewed journal papers and over 100 conference papers. Prof. Gustafsson received the IEEE Schelkunoff Transactions Prize Paper Award 2010 and Best Paper Awards at EuCAP 2007 and 2013. He served as an IEEE AP-S Distinguished Lecturer for 2013-15.
\end{biography}

\begin{biography}[{\includegraphics[width=1in,height=1.25in,clip,keepaspectratio]{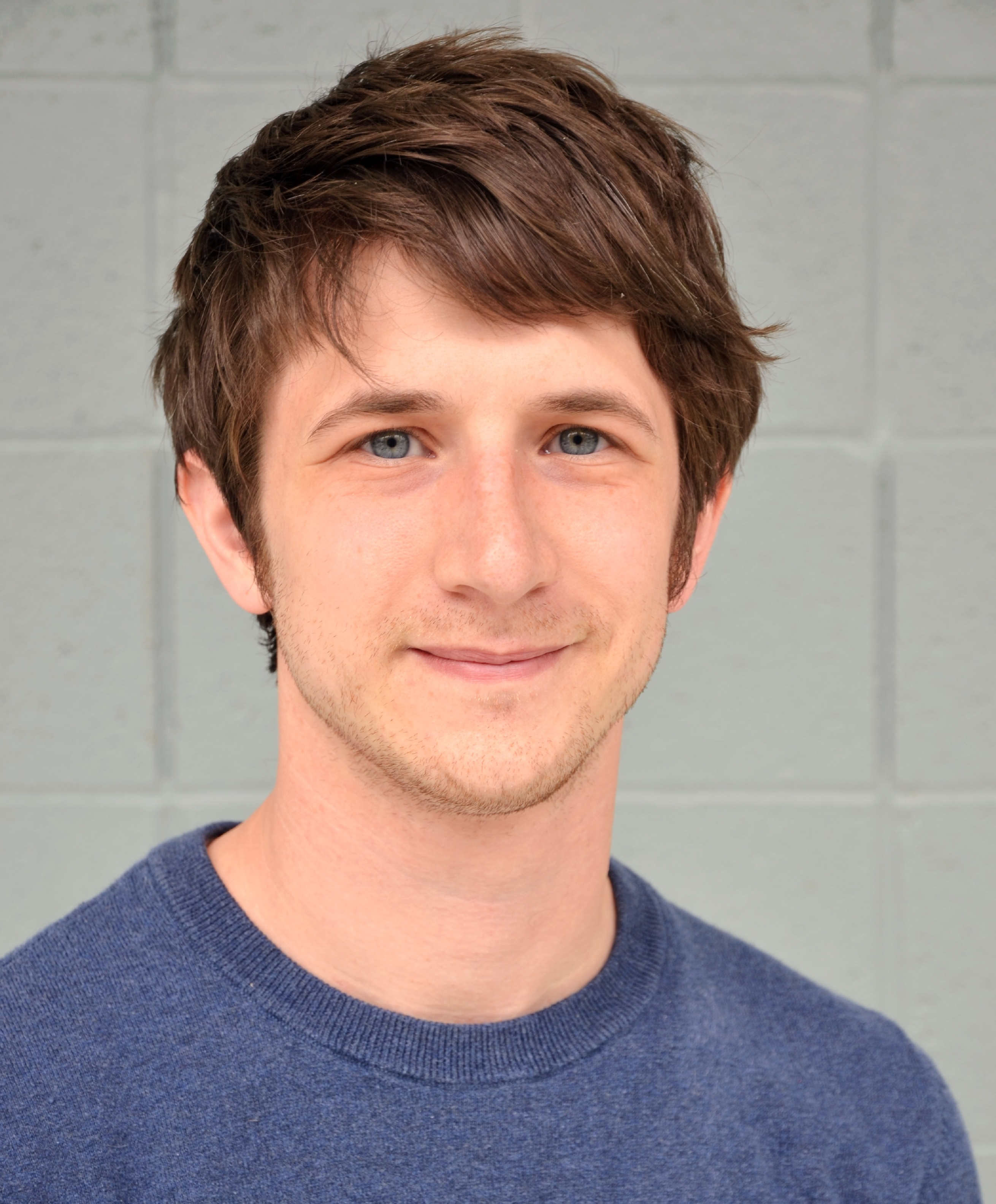}}]{Kurt Schab}
(S'09) received the B.S. degree in electrical engineering and physics from Portland State University, Portland, OR, USA, in 2011, and the M.S. and Ph.D. degrees in electrical engineering from
the University of Illinois at Urbana-Champaign, Champaign, IL, USA, in 2013 and 2016, respectively.  Currently, he is a postdoctoral research fellow at North Carolina State University.

His research interests include electromagnetic scattering theory, optimized antenna design, and numerical methods in electromagnetics.
\end{biography}
\end{document}